\documentclass[prl,twocolumn,showpacs,byrevtex]{revtex4}
\input epsf
\usepackage{graphicx}
\usepackage{dcolumn}
\usepackage{bm}

\def\dsodt{ds_{23}\over dt}
\def\dstdt{ds_{13}\over dt}
\def\dsthdt{ds_{12}\over dt}
\def\uoo{c_{13}c_{12}}
\def\uot{c_{13}s_{12}}
\def\uoth{s_{13}}
\def\uto{-c_{23}s_{12}-c_{12}s_{13}s_{23}}
\def\utt{c_{12}c_{23}-s_{12}s_{13}s_{23}}
\def\utth{c_{13}s_{23}}
\def\utho{s_{12}s_{23}-c_{12}s_{13}c_{23}}
\def\utht{-c_{12}s_{23}-c_{23}s_{13}s_{12}}
\def\uthth{c_{13}c_{23}}
\def\be{\begin{equation}}
\def\ee{\end{equation}}
\def\ba{\begin{eqnarray}}
\def\ea{\end{eqnarray}}
\def\br{\begin{array}}
\def\er{\end{array}}

\def\Dmott{\Delta m^2_{12}}
\def\Dmttht{\Delta m^2_{23}}

\begin{document}
\title{ High Scale Mixing Unification and Large Neutrino Mixing Angles}
\author{R. N. Mohapatra}
\email{rmohapat@physics.umd.edu}
\affiliation{Department of Physics, University of Maryland,
College Park, MD 20742, USA.}
\author{M. K. Parida}
\email{mparida@sancharnet.in}
\affiliation{Department of Physics, North Eastern Hill University,
Shillong 793022, India.}
\author{G. Rajasekaran}
\email{graj@imsc.ernet.in}
\affiliation{Institute of Mathematical Sciences, Chennai 600 113, India.}
\begin{abstract}
Starting with the hypothesis that quark and lepton mixings are
identical at or near the GUT scale, we show that the large solar and
atmospheric neutrino mixing angles together with the small reactor angle
$U_{e3}$
can be understood purely as a result of renormalization group evolution
provided the three neutrinos are quasi-degenerate and have same CP
parity. It predicts the common Majorana mass for the neutrinos larger than
$0.1$ eV, which falls right in the range reported recently and also the
range which will be probed in the planned experiments.
\end{abstract}
\date{\today}
\rightline{UMD-PP-03-038}
\rightline{NEHU/PHY-MP-02/03}
\rightline{IMSc/2003/01/02}
\maketitle
\section{I. INTRODUCTION}
\label{sec1}
The idea that disparate physical parameters describing forces and matter
at low
energies may unify at very short distances (or high mass scales) has been
a very helpful tool in seeking a unified understanding of apparently
unrelated phenomena \cite{ref1}. In the context of supersymmetric grand
unifed theories, such an approach explains the weak mixing angle
$\sin^2\theta_W$ and thereby the different strengths of the weak,
electromagnetic and strong forces. One of the key ingredients of the grand
unified theories is the unification between quarks and leptons. One may,
therefore, hope that in a quark-lepton unified theory, the weak interaction
properties of quarks and leptons parameterized by means of the flavor
mixing matrices will become identical at high energies. 

On the
experimental side, recent measurements on atmospheric and solar neutrino
fluxes and those at K2K and KamLAND which are a manifestation of the
phenomena of
neutrino oscillations suggest that two of the neutrino mixings i.e.
the mixings between
$\nu_e- \nu_{\mu}$ and  $\nu_{\mu}-\nu_{\tau}$
(to be denoted by $\theta_{12}$ and $\theta_{23}$, respectively)
are large \cite{ref2,ref3,ref4,ref5, ref6} while the third mixing
between the $\nu_e-\nu_{\tau}$ is bounded to be very small by the
CHOOZ-Palo Verde reactor
experiments i.e.  $\sin^2{2\theta_{13}}< 0.15$
\cite{ref7}. On the other hand, it is now quite well established that
all observed quark mixing angles are very small. One may therefore ask
whether there is any trace of quark lepton unification in the mixing
angles as we move to higher scales.

The first question in this connection is whether high scales have anything
to do with neutrino masses or it is purely a weak scale phenomenon. One of
the simplest ways to understand small neutrino masses is via the
seesaw mechanism \cite{ref8} according to which the neutrino mixing
is indeed a
high scale phenomenon, the new high scale being that of
 the right handed neutrino masses ($M_R$) in an appropriate extension of
the standard model. Present data put the seesaw
scale $M_R$ very close to the conventional GUT scales. It is therefore
tempting to speculate whether quark and lepton mixing angles are indeed
unified at the GUT-seesaw scale. This would of course imply that 
all neutrino mixing angles at the high scale $M_R$ are very small whereas
at the weak scale two of them are known to be large. In this paper we
show that simple radiative correction effects embodied in the
renormalization group evolution of parameters from seesaw scale to the
weak scale can indeed provide a complete understanding of all neutrino
mixings at the weak scale, starting with very small mixings at the
GUT-seesaw scale. 

\par
The fact that renormalization group evolution from the seesaw
scale to the weak scale \cite{ref9,ref10} can lead to drastic changes
in the magnitudes of the mixing angles was
pointed out in several papers \cite{ref9, ref11, ref12, ref13, ref14, 
ref16}.
In particular, it was shown in \cite{ref11} that this dependence on
renormalization group
evolution can be exploited in simple seesaw extensions of the minimal
supersymmetric standard model (MSSM) to explain the large value of the
atmospheric mixing angle starting with a small mixing at the seesaw scale,
provided two conditions are satisfied: (i) the two neutrino-mass eigen
states have same CP-parity  and (ii) they are very nearly degenerate 
in mass. In
general, in gauge models that attempt to explain the large neutrino
mixings \cite{ref17}, one needs to make many assumptions to constrain the
parameters. In contrast, in this class of ``radiative magnification''
models \cite{ref11, ref12, ref14},~there is no need to invoke
special constraints on the parameters at high scales  beyond those needed
to guarantee the quasi-degeneracy. In fact  the main content of radiative 
magnification models is the quasi-degeneracy assumption and since the  
value of common
Majorana mass $m_0$ for all neutrinos
is required 
to be in sub-eV range ($\geq 0.1$ eV,), this assumption is 
experimentally testable in the ongoing neutrinoless double beta decay
searches \cite{ref12}. 

It is well known that the radiative 
magnification technique requires adjustment of initial neutrino mass eigen
values at the see-saw scale \cite{ref9, ref11, ref12, ref13, ref14}. In  
view of the model
independence and simplicity of the method involved, and the attractive nature 
of the 
results achieved, the question of finetuning has been discussed at lengh by 
Casas, Espinosa, Ibarra, and Navarro(hereafter called as CEIN) \cite{ref14}
who have also  discussed  the relevant  magnification criteria 
and shown that, in three flavor case, the existence of infrared stable 
quasifixed
points in the relevant RGEs lead to vanishing mixing matrix elements at
low energies. Thus, magnification for mixing angles is expected to occur 
only if RG-evolutions are stopped before reaching the quasifixed point regime.
It has been noted that the radiative magnification mechanism leading to large 
neutrino mixing can only be achieved   
if there is substantial cancellation between the initial and the
RG-generated mass splittings \cite{ref11, ref12, ref13, ref14}. In this 
context we note 
that similar cancellations are common in well known grand unified 
theories(GUTs). In the bottom-up approach large differences between 
low-energy coupling constants of the SM are reduced to vanishing 
differences due to cancellation with RG-generated contributions. 
In the well known $b-\tau$ unification scenario, the low-energy mass
splitting ${\rm m}_b-{\rm m}_{\tau} \simeq 2.5$ GeV is almost cancelled out 
by RG-generated mass difference leading to ${\rm m}_b \simeq {\rm m}_{\tau}$
at the GUT scale. Both these examples apply to nonSUSY as well as SUSY GUTs.
 
In this paper\cite{ref15}, we show that under the same
conditions for radiative magnification as just outlined, if we start with
the hypothesis that at the
seesaw scale the quark and neutrino mixings are unified to a common set of
values, i.e. the known extrapolated values of the well known CKM angles, 
after renormalization group evolution to the weak scale, we can obtain the
solar and the atmospheric mixing angles that are in agreement with
observations  without contradicting the CHOOZ-Palo Verde
bound on $\theta_{13}$. The possibility of achieving two large neutrino 
mixings by radiative magnification has been  reported for the first time  
in ref.\cite{ref15}.  
 
This result has two important implications: (i) it would provide a very
simple and testable way to understand the observed large neutrino mixings
and (ii) if confirmed by the neutrinoless double beta decay experiments,
it would provide a strong hint of quark lepton unification at high scales. 
One may wonder why we are addressing the question of unification of
the mixing angles for neutrinos with those of quarks and not the
unification of neutrino masses with quark masses. The answer is of
course the well-known one, namely neutrino masses have their origin
(seesaw mechanism) that distinguishes them from the quark
masses. Furthermore, within the seesaw mechanism neutrinjos are Majorana
fermions whereas the quarks are Dirac fermions. Thus as far as the masses
go, we have no reason to expect unification with quarks. We take up the
question of neutrino masses in Sec 5.

This paper is organized as follows: in sec. 2, we discuss the RGes for the
neutrinos in the mass basis, in sec. 3, we present the main result of our
paper i.e. the magnification of mixing angles at the weak scale; in
sec. 4, we discuss predictions of our approach for neutrinoless double
beta decay and other processes; in sec. 5, we present a gauge model where
approximate mixing unification hypothesis is realized and in sec. 6, we
present our conclusions.

\par
\section{II. RENORMALIZATION GROUP EQUATIONS FOR MASSES AND  MIXINGS}
\label{sec2}
Our basic assumption will be a seesaw type model which will lead to equal
quark and lepton mixing angles at the seesaw scale as well as to a
quasi-degenerate neutrino spectrum. In sec. 5, we present a model where at
the seesaw scale the neutrinos have this property. We will
then follow the ``diagonalize and run'' procedure for the neutrino
parameters and use the RGEs directly for the physical observables,
namely, the mass eigenvalues $m_i$ and the mixing angles $\theta_{ij}$
($i,j=1,2,3$).  We also assume  the
neutrino mass eigenstates to possess the same CP and ignore CP
violating phases in the mixing matrix. Also for
 simplicity, we adopt the mass ordering among the
quasi-degenerate eigenstates to be of type $m_3\agt m_2\agt
 m_1$. The real $3\times 3$ mixing matrix is
parametrized as,
\par
\noindent
\be U=\left[\br{ccc}
\uoo&\uot&\uoth\\
\uto&\utt&\utth\\
\utho&\utht&\uthth
\er\right],\label{eq1}\ee
\par
\noindent
where $c_{ij}=\cos\theta_{ij}$ and $s_{ij}=\sin\theta_{ij} (i,j=1, 2,
3$). $U$
diagonalizes the mass matrix
$M$ in the flavor basis with $U^TMU={\rm diag}(m_1, m_2, m_3)$.
The RGEs for the mass eigen values  can be written as \cite{ref13, ref14}
\par
\noindent
\be{dm_i\over dt}=-2F_{\tau}m_iU_{\tau
i}^2-m_iF_u,\,\left(i=1,2,3\right).\label{eq2}\ee
\par
\noindent
For every $\sin\theta_{ij}=s_{ij}$, the corresponding RGEs are,
\par
\noindent
\ba\dsodt&=&-F_{\tau}{c_{23}}^2\left( 
-s_{12}U_{\tau1}D_{31}+c_{12}U_{\tau2}D_{32}
\right),\label{eq3}\\
\dstdt&=&-F_{\tau}c_{23}{c_{13}}^2\left( 
c_{12}U_{\tau1}D_{31}+s_{12}U_{\tau2}D_{32}
\right),\label{eq4}\\
\dsthdt&=&-F_{\tau}c_{12}\left(c_{23}s_{13}s_{12}U_{\tau1}
D_{31}-c_{23}s_{13}c_{12}U_{\tau2}D_{32}\right.\nonumber \\
&&\left.+U_{\tau1}U_{\tau2}D_{21}\right).\label{eq5}\ea
\par
\noindent
where $D_{ij}={\left(m_i+m_j)\right)/\left(m_i-m_j\right)}$
and,~for MSSM,
\par
\noindent
\ba F_{\tau}&=&{-h_\tau^2}/{\left(16\pi^2\cos^2\beta\right)},\nonumber\\
F_u&=&\left(1\over{16\pi^2}\right)\left({6\over5}g_1^2+6g_2^2-
6{h_t^2\over\sin^2\beta}\right),\label{eq6}\ea
\par
\noindent
but, for SM,
\par
\noindent
\ba F_{\tau}&=&{3h_\tau^2}/\left( 32\pi^2\right),\nonumber \\
F_u&=&\left(3g_2^2-2\lambda-6h_t^2-2h_\tau^2\right)/\left(16\pi^2\right).
\label{eq7}\ea
\par
\noindent
RGEs of mixing angles in  the three flavor case have been shown to possess
infrared stable quasifixed points leading to  vanishing values of
mixing matrix elements \cite{ref14}. Thus as in two-flavor case 
\cite{ref11,ref12}
the radiative magnification of two mixing angles, if at all feasible,
could be realizable  only if RG-evolution is stopped before 
reaching the quasifixed point regime. In the CEIN \cite{ref14} approach 
this was implemented to magnify the atmospheric mixing angle 
by adjusting the initial mass eigen values to achieve maximal mixing at
$M_{\rm SUSY}\sim M_Z$ such that decrease in mixing angles after reaching 
the maximum is smaller. In this approach  MSSM operates for all scales 
starting from  $M_{Z}$. 

In order to avoid the vanishing matrix elements of the quasifixed point 
region in this paper we follow a different approach described in \cite{ref12}
which is found to work also for three flavor case in the presence of 
MSSM with $O$(TeV) SUSY scale.When 
the mass difference between $m_i$ and $m_j$ tends to vanish,~$D_{ij}\to
\infty$, and the corresponding term in the RHS of (\ref{eq3})-(\ref{eq5})
predominantly drives the RG evolution for  $\sin\theta_{ij}$
which might become large or even approach its  maximal
value anywhere between
$\mu=M_{\rm SUSY}-M_R$. Since $m_i$ and $m_j$ are scale dependent, the
initial difference existing between them at $\mu=M_R$ is narrowed down
during
the course of RG evolution as we approach $\mu=M_{\rm SUSY}$. This causes
$D_{ij}\to\infty$ and hence large magnification to the mixing angle due to
radiative effects. Also $F_\tau$ is enhanced by a factor $\sim 10^3$ in the
large $\tan\beta$ region in the case of MSSM
as compared to the SM where such effects do not exist. Thus, if the SUSY scale
is significantly larger  with $M_Z < M_{\rm SUSY} \le 1$TeV, radiative
magnification to large mixings may occur through RG evolution from the see-saw
scale  down to $M_{\rm SUSY}$. Then the standard model
evolution  below $M_{\rm SUSY}$
causes negligible contribution to  the magnified mixings
because of two reasons:
(i)absence of $\tan^2\beta$ effects, and (ii)small range of RG evolution
from $M_{\rm SUSY}$ to $M_Z$. Consequently, the predicted mixings remains 
almost flat
and very close to $\sin\theta_{ij}(M_{\rm SUSY}$) for all energies below
$M_{\rm SUSY}$.
This aspect of RG-evolution below the SUSY scale to avoid the approach
to infrared stable quasifixed point corresponding to vanishing mixing 
angle which  was  demonstrated in 
\cite{ref12} is also found to  operate in three flavor case. .
\par
The mixing unification hypothesis implies that we choose all neutrino
mixings at the seesaw scale equal to the corresponding quark mixings,
which in the Wolfenstein parameterization are dictated by the parameter 
$\lambda_0=.2$. We then have, at the seesaw scale,
 $s_{12}\simeq\lambda_0$, $s_{23}\simeq O(\lambda_0^2)$ and
$s_{13}\simeq O(\lambda_0^3)$. These values get substantially magnified
in the region around $M_{\rm SUSY}$. Using 
$|D_{31}|\simeq |D_{32}|\ll |D_{21}|$, we see from 
(\ref{eq3})-(\ref{eq5}), that the dominant contribution to RG evolution of
$s_{23}(\mu)$ is due
to the term $\sim \lambda_0^2F_{\tau}D_{32}$. Similarly the terms 
contributing
to the evolution of $s_{13}(\mu)$ are $\sim \lambda_0^3F_{\tau}D_{32}$ or
$\sim\lambda_0^3F_{\tau}D_{31}$. On the other hand 
the evolution of $s_{12}$ is dominated 
by the term $\sim \lambda_0^5F_{\tau}D_{21}$ where the large enhancement
likely to be caused by the
largeness in $|D_{21}|$ is damped out due to higher 
power of $\lambda_0^5$.
Since the mixing angles change substantially 
only around $M_{\rm SUSY}$, such
dominance to RG evolutions holds approximately at 
all other lower scales below $M_R$. 

\par
If the neutrino mixing angles are to be compatible with experimental
observations at low energies, we need at most the  magnification
factors:
${(\sin\theta_{23}/\sin\theta_{23}^0)}\simeq 20$,
${(\sin\theta_{13}/\sin\theta_{13}^0)}
\le 60$, and
 ${(\sin\theta_{12}/\sin\theta_{12}^0)} \simeq 4$,
where we have used the experimental neutrino mixings for $\theta_{ij}$
\cite{ref2, ref3, ref4, ref5, ref6, ref7} and quark mixings for
$\theta_{ij}^0$ \cite{ref18}.
~That the CHOOZ-Palo Verde bound can tolerate  a magnification factor as large
as 60 is crucial to achieve bi-large mixings by radiative magnification while
keeping the magnified angle $\theta_{13}$ at low energies well below the upper
bound. This is of course because of the smallness of $(\lambda_0)^3$,
which is
the starting value (order of magnitude) of the reactor angle. One
can also observe that it is the smallness of the reactor angle that
provides the "hidden" signal for the unification!

\section{III. BI-LARGE NEUTRINO MIXINGS BY RG EVOLUTION}
\label{sec3}

Starting from known values of gauge couplings,  masses of quarks and charged
leptons, and CKM mixings in the quark sector at low energies, at first we
use the bottom-up approach and all the relevant RGEs to obtain the
corresponding quantities at higher scales,
$10^{11}$ GeV-$2\times 10^{18}$ GeV. Assuming the neutrino mixing at
$\mu=M_R$
to be small and similar to quark mixings, we then expect the initial
conditions at $\mu=M_R$ to be $\sin\theta^0_{23}\simeq 0.038$, 
$\sin\theta^0_{13}\simeq 0.0025$
and $\sin\theta^0_{12}\simeq 0.22$ \cite{ref18}. Using these as input and the
mass eigenvalues $m^0_i$ as unknown parameters at the high scale, we then
follow  the top-down approach though
(\ref{eq2})-(\ref{eq5}) and other standard RGEs. The unknown
parameters $m^0_i$
are determined in such a way that the solutions
obtained at low energies
agree with mass squared differences and the mixing angles given by the
experimental data within $90\%$ C.L.~\cite{ref2, ref3, ref4, ref5, 
ref6, ref7}
\par
\noindent
\ba \Dmott&=&\left(2-50\right)\times 10^{-5}{\rm eV}^2,\nonumber \\
    \Dmttht&=&\left(1.2-5\right )\times10^{-3}{\rm eV}^2,\nonumber \\
    \sin\theta_{23}&=& 0.54-0.83, \sin\theta_{12}= 0.40-0.70, \nonumber \\
    \sin\theta_{13}&\le& 0.16~.\label{eq8}\ea
\par
Our model described in sec.5 is consistent with quasidegenerate mass eigen 
values over a wider range of the see-saw scale: $M_R= 10^{11}$ GeV-$10^{15}$ 
GeV. However, in view of the phenomenological importance of the rersults, 
we have explored the RG-evolutions to bi-large mixings including  
higher scales upto the reduced Planck scale($2\times 10^{18}$ GeV). 
In Table I we prersent input mass eigenvalues at the see-saw scale 
$M_R=10^{13}$ GeV and
our solutions  at  $M_Z$ in the large $\tan\beta$($=55$) region. 
The solutions clearly exhibit radiative magnification of
both the mixing angles, $\theta_{23}$ and $\theta_{12}$
for a wide range of input values of $m_i^0$. We find that
although
enhancement due to RG evolution occurs in the $\nu_e-\nu_{\tau}$
sector also,  $\sin\theta_{13}$ remains
well within the CHOOZ-Palo Verde bound \cite{ref7}. 

In Table II we present
three sets of initial mass eigenvalues and our solutions 
for three different high-scale values, $M_R= 10^{11}, 10^{15}$ and 
$2\times 10^{18}$ GeV. We find that
for the same value of $\tan\beta=55$, the predicted lowest  mass eigenvalue
at $M_Z$ decreases slowly with increase of the  see-saw scale. For example, 
the lowest mass eigenvalues predicted at $\mu =M_Z$ are $0.27$ eV, $0.22$ eV
, $0.209$ eV, and $0.17$ eV for $M_R$= $10^{11}$ GeV, $10^{13}$ GeV,
$10^{15}$ GeV and $2\times 10^{18}$ GeV , respectively. 

A magnification  formula has been  derived by CEIN \cite{ref14} for
the product of the mixing matrix elements,
\par\noindent
\ba
F_{\rm m}&=& {U_{\tau i}U_{\tau j}(\mu) \over U_{\tau i}U_{\tau j}(M_R)}
\nonumber\\
&\simeq& \left[1+{h_{\tau}^2\over {32\pi^2}}D_{ij}(M_R)\ln{M_R\over
\mu}\right]^{-1}.\label{eq9}\ea
Using the values given in Tables I-II, we find that the magnification 
predicted by the formula matches reasonably well with our estimations for 
mixing 
between the second and the third generations(i, j=2, 3). 	        
\par
\begin{figure}
\epsfxsize=8.5cm
\epsfbox{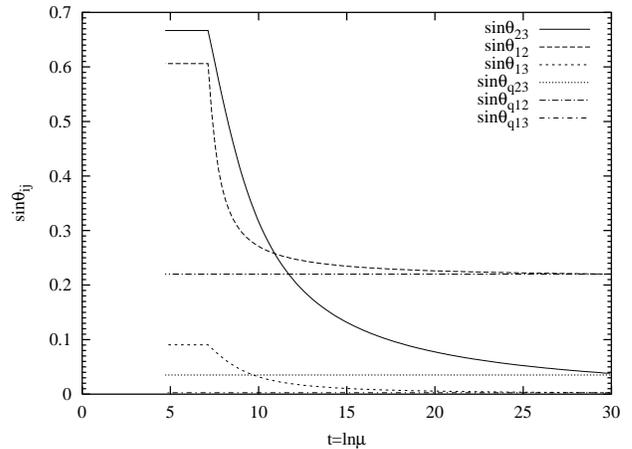}
\caption{Evolution of small quark-like mixings at the see-saw scale to bilarge
neutrino mixings at low energies for the see-saw scale $M_R=10^{13}$ GeV with
input and output mass-eigen values and mixing angles given in the first column
of Table I. The solid, long-dashed and short-dashed 
lines represent
$\sin\theta_{23}$, $\sin\theta_{13}$, and $\sin\theta_{12}$, respectively, 
as defined in the text. Almost horizontal lines represent the corresponding
sines of the CKM mixing angles in the quark sector}  
\label{fig1}
\end{figure}

Our result on the approximate unification of quark 
and neutrino mixings at the high scale $M_R=10^{13}$ GeV is exhibited in 
Fig.\ref{fig1} where
 we present the RG evolutions of the sines of
the three neutrino mixing angles
starting from $M_R=10^{13}$ GeV down to $M_Z$ for one  set of
input masses given in Table I:
$m^0_1=0.2983$ eV, $m^0_2=0.2997$ eV, and $m^0_3=0.3383$ eV.
The flatness of the curves below $M_{SUSY}$ is due to negligible
renormalization effect from SM which evades the approach to 
the quasifixed points.
 The corresponding low-energy solutions are
$m_1=0.2201$ eV, $m_2=0.2223$ eV,  $m_3=0.2244$ eV,
~$\Dmott=1.6\times10^{-4}$ eV$^2$, $\Dmttht=1.0\times 10^{-3}$ eV$^2$,
~$\sin\theta_{23}=0.667$, $\sin\theta_{13}=0.09$, and
$\sin\theta_{12}=0.606$.
Almost horoizontal lines in the figure represent  the sines of the CKM mixings, 
$\sin\theta^q_{ij}$,   having negligible one-loop radiative corrections 
\cite{ref18}. 
Unification of the 
neutrino mixings with the corresponding quark mixings are clearly 
demonstrated at the high scale.

\begin{figure}
\epsfxsize=8.5cm
\epsfbox{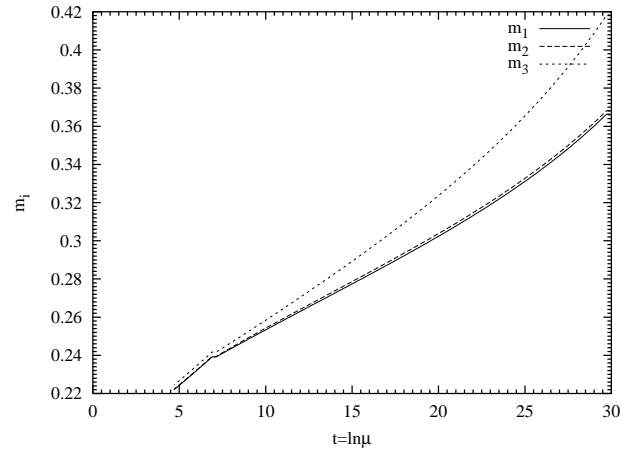}
\caption{Evolution of neutrino mass eigenvalues from the see-saw scale 
$M_R=10^{13}$ GeV  corresponding to initial values $m_1^0=0.3682$ eV, $m_2^0=0.37$ eV, and $m_3^0=0.4210$ eV, leading to the predictions  
$m_1=0.2201$ eV, $m_2^0=0.2223$ eV, and $m_3^0=0.2244$ eV at $M_Z$ and bi-largemixings as shown in Fig.1 and Table I}                          
\label{fig2} 
\end{figure}

The evolution of mass eigen values corresponding to mixings given in 
Fig. \ref{fig1} are shown in Fig.\ref{fig2} for $M_R=10^{13}$ GeV. In 
cntrast to sines of mixing angles  which have negligible RG-corrections 
below the SUSY scale, the mass eigen values are found to decrease till the 
lowest scale $M_Z$. The rate of
decrease of the third eigen value is the highest, but the rates of decrease of the first and the second eigen values are similar. The initial mass splittings at the higest scale are narrowed down to match the experimental values at low energies due to cancellations caused by RG-generated splittings.     

\begin{figure}
\epsfxsize=8.5cm
\epsfbox{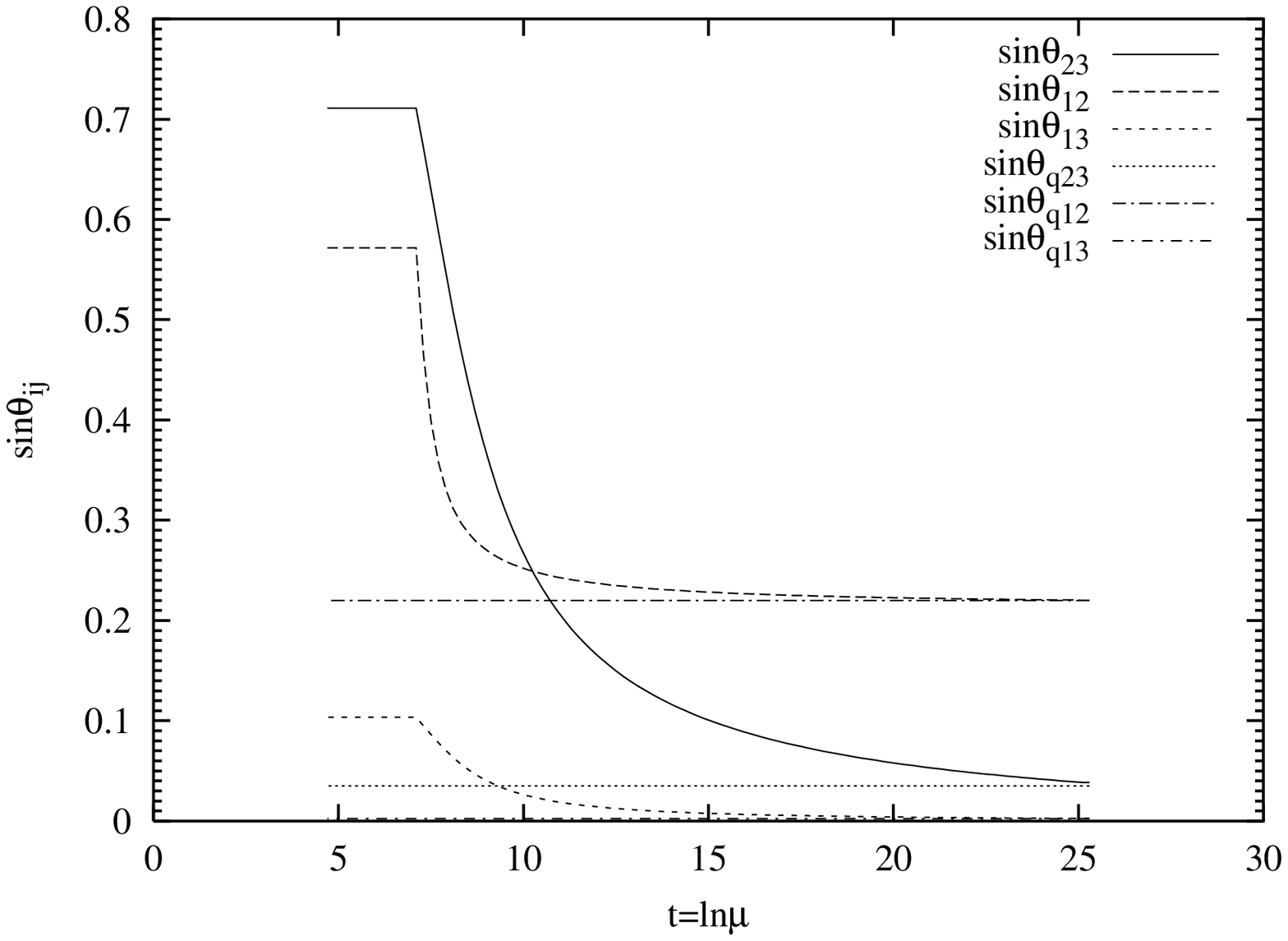}
\caption{Same as Fig.1 but for $M_R=10^{11}$ GeV and inputs given in Table II} 
\label{fig3}
\end{figure}

\begin{figure}
\epsfxsize=8.5cm
\epsfbox{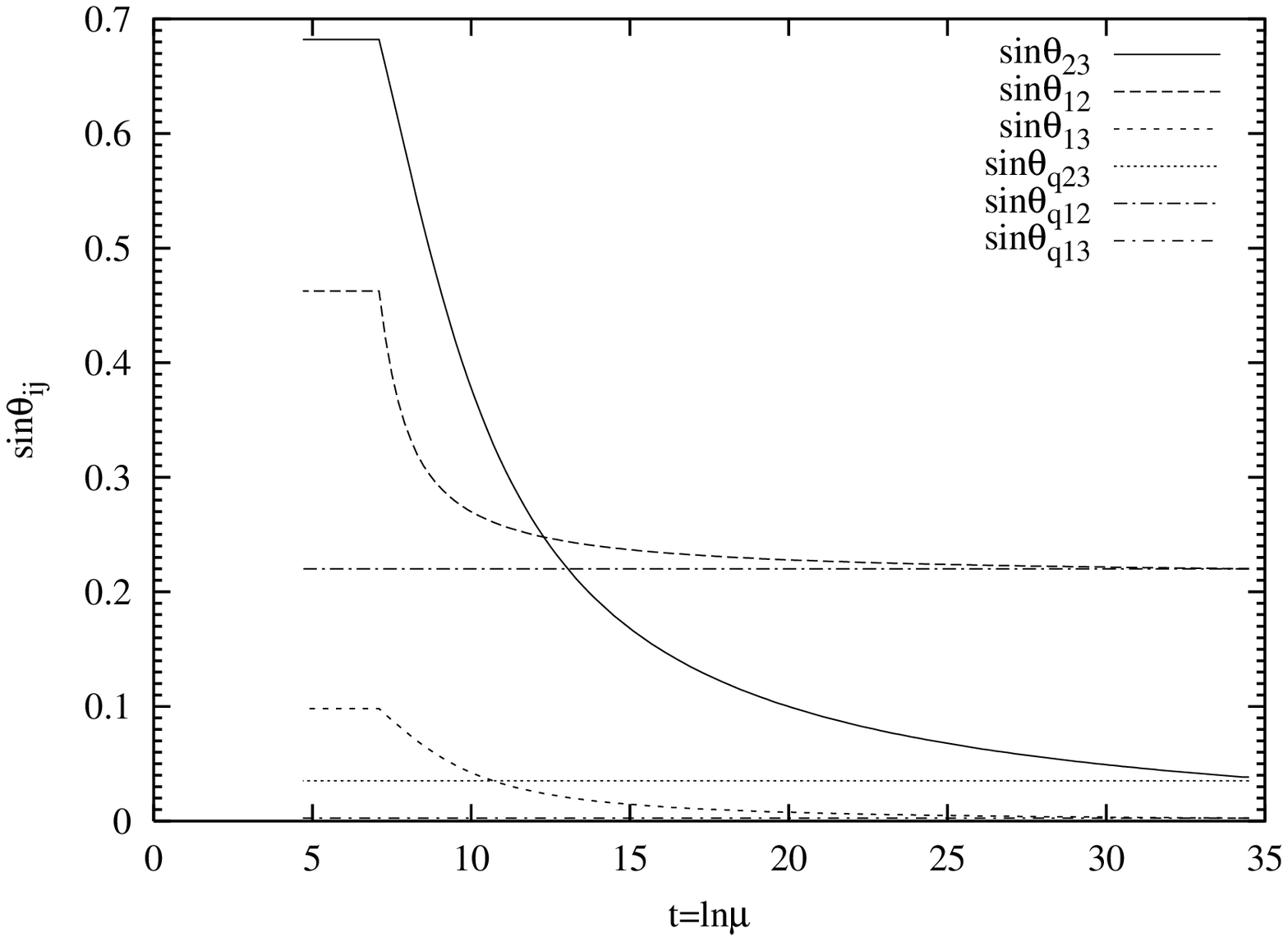}
\caption{Same as Fig.1 but for $M_R=10^{15}$ GeV and input given in Table II  }\label{fig4} 
\end{figure}

\begin{figure}
\epsfxsize=8.5cm
\epsfbox{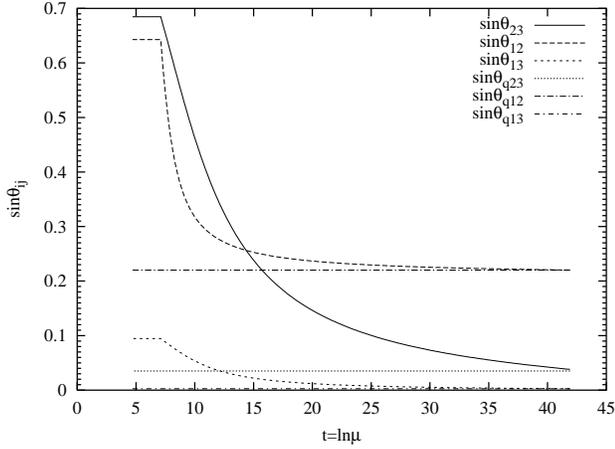}
\caption{Same as Fig.1 but for  $M_R=2\times 10^{18}$ GeV and input given in Table II}  
\label{fig5}
\end{figure}

\begin{figure}
\epsfxsize=8.5cm
\epsfbox{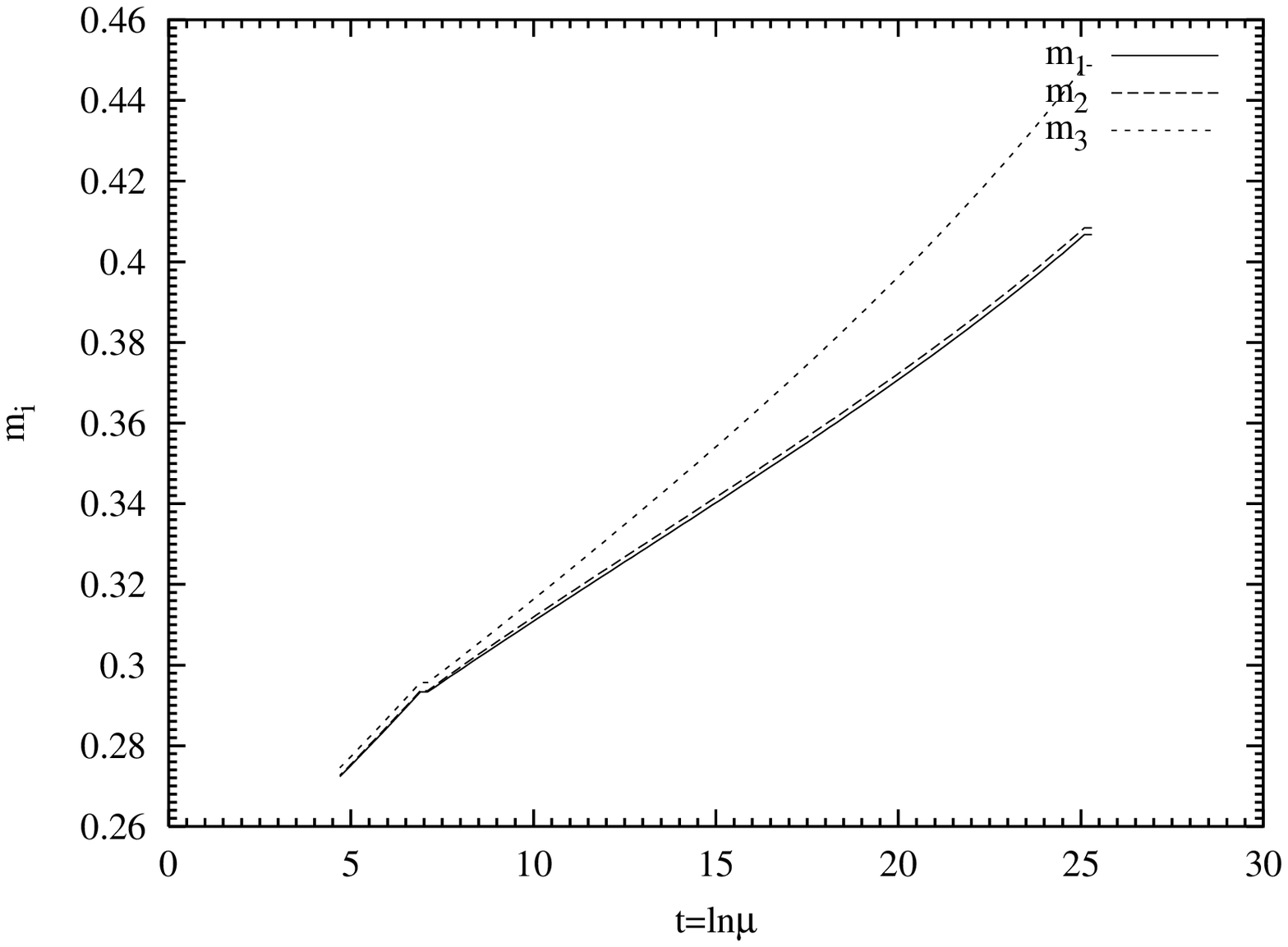}
\caption{Same as Fig.2 but for $M_R=10^{11}$ GeV and input given in Table II}  \label{fig6} 
\end{figure}

\begin{figure}
\epsfxsize=8.5cm
\epsfbox{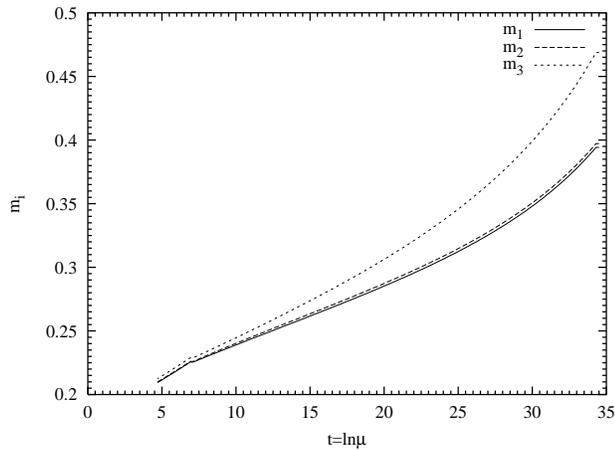}
\caption{Same as Fig.2 but for $M_R=10^{15}$ GeV and input values given in Table II}  
\label{fig7}
\end{figure}

\begin{figure}
\epsfxsize=8.5cm
\epsfbox{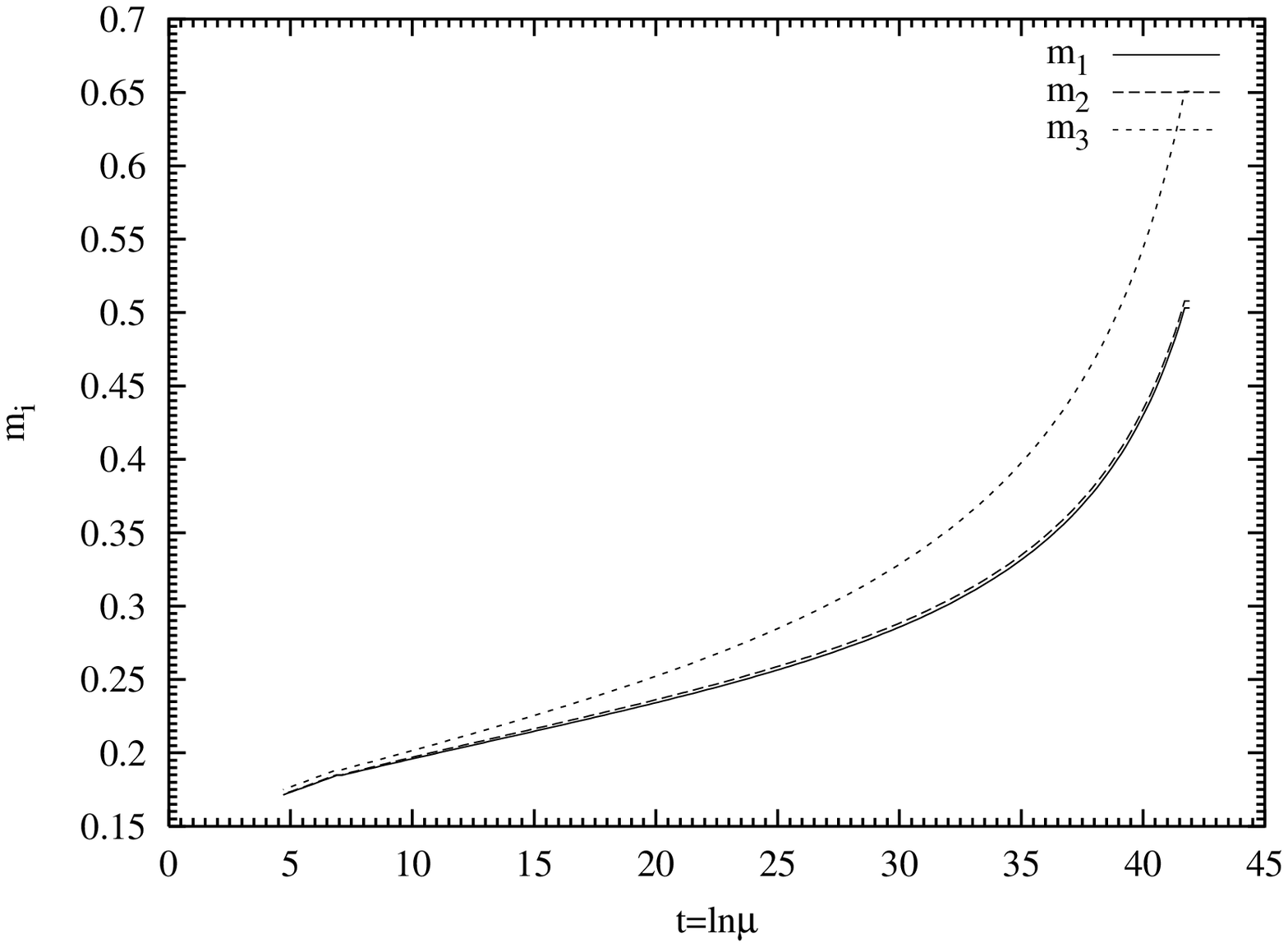}
\caption{Same as Fig.2 but for $M_R=2\times 10^{18}$ GeV and injput values given in Table II }                          
\label{fig8} 
\end{figure}

In Fig.\ref{fig3}-Fig.\ref{fig5} we present evolutions of neutrino mixing angles for
$M_R=10^{11}$ GeV, $10^{15}GeV$ and $2\times 10^{18}$ GeV with input mass 
eigen values given in Table II. 
In Fig.\ref{fig6}-Fig.\ref{fig8} the RG-evolution of corresponding mass eigen 
values  with input parameters given in Table II
for $M_R=10^{11}$ GeV, $10^{15}$ GeV, and $2\times 10^{18}$ GeV are 
presented.
It is quite clear that radiative magnification
to bi-large mixings is possible over a wide range of choices of $M_R$
and input mass eigenvalues.

In addition to the solutions of the type shown in Fig. \ref{fig1} which are 
valid
for $M_{\rm SUSY}>M_Z$, we have also found solutions corresponding to
two large and one small mixings  for $M_{\rm SUSY}=M_Z$ with somewhat
different  mass eigen values in agreement with the 
experimental data at low energies. 
We have also noted that the radiative magnification mechanism leading 
to bilarge
mixings works more easily if we take all other initial values same  
as mentioned above but 
$\sin\theta^0_{13}=0.0$ which could be relevant to certain neutrino 
mass textures. In
this case the CHOOZ-Palo Verde bound is  always protected.

It is worth re-emphasizing that since we determine 3 input parameters (the
3 mass eigenvalues at
high scale) to fit 5 experimentally known numbers as output parameters
it is a over-determined problem and there may be no solution. So
there is a possibility of not being able to obtain correct mixing
angles at the weak scale. But we have found that it is possible, thus
showing that there is perhaps an
element of truth in the unification hypothesis. It is also significant
that the scale of 0.16 - 0.65 eV comes out as the range of allowed
mass eigenvalues although such a scale was not put in at all a priori.

\par
\section{IV. PREDICTIONS FOR BETA-DECAY, DOUBLE BETA DECAY, $U_{e3}$ AND
WMAP}
\label{sec4}
  
	Very recently the possibility of testifying our mixing unification
hypothesis through lepton-flavor violating processes like $\mu \to {\rm e}
\gamma$ and $\tau \to \mu \gamma$ has been investigated  \cite{ref19}.
 We discuss here other     
 possible  experimental tests of the specific mechanism
proposed here for radiative magnification.

\noindent{\bf Double beta and tritium beta decays: }  Our RG solutions
permit
radiative magnification consistent with experimental data on $\Delta
m^2_{21}$, $\Delta m^2_{32}$ and the mixing angles, if the input mass
eigenvalues
for $M_R = 10^{11}-2\times 10^{18}$ GeV are in the range 0.35 eV - 1.0 eV. 
This corresponds
to the low energy limits $0.16 ~eV < m_i(M_Z) < 0.65 ~eV$. Then,our choice
of phases leads to the prediction
\begin{eqnarray}
    |<M_{ee}>|=|\sigma_i m_i U^2_{ei}|= 0.16 ~eV -0.65 ~eV.     
\end{eqnarray}
Recent searches for neutrinoless double beta decay have obtained the
upper limit: $|<M_{ee}>| < (0.33 - 1.35)$ ~eV \cite{ref20}. The range in Eq (8)
overlaps the one reported in \cite{ref21}, or the ones that will be covered
in \cite{ref22}. Thus a clear and testable prediction of the bi-large radiative
magnification mechanism is that neutrinoless double beta decay should
be observed in the next round of experiments.

  Further, our low-energy limit on the quasi-degenerate $m_i(M_Z)$ can be
directly measured in Tritium beta decay experiment. Although the present
experimental bound on the mass is $< 2.2$ eV, mass value as low as 0.35 eV
can be reached by KATRIN experiment \cite{ref23}.

\par\noindent {\bf Prediction for $U_{e3}$:}
Starting from the  allowed range of high-scale input values of  the CKM
mixing angle with  $V_{ub}\simeq U^0_{e3}\simeq 0.0025-0.004$, the
RG-evolutions predict enhancement of $\sin\theta_{13}$ at low
energies
\be U_{e3}=\sin\theta_{13}=0.08-0.10.  \ee
Although this prediction is well below the present experimental upper
bound \cite{ref6}, it is accessible to several planned long-baseline
neutrino experiments in future such as NUMI-Off-Axis or JHF proposals.

\par\noindent{\bf WMAP constraints on neutrino masses:}
Recently the Wilkinson Microwave Anisotropy Probe (WMAP) observations have
provided very interesting constraints on the sum of neutrino
masses \cite{ref24, ref25}. The analysis depends on a number of cosmological
parameters such as $H_0$, the bias parameter $b(k)$, $\Omega_m$ from SN-Ia
observations etc. Depending on what values one chooses for the ``priors'',
the constraint on the the sum all neutrino masses varies from $2.1$ eV to
$0.7$ eV. Since we are proposing that the neutrino masses are degenerate,
each individual mass will have an upper limit of $0.23$ eV to $0.7$ eV.
Thus the radiative magnification hypothesis is  consistent with
WMAP observations \cite{ref24} and also with the combined analysis of 
WMAP$+$2dF GRS data \cite{ref25}.

  We have found that with $\tan\beta=55$ and due to RG-effects alone
the lowest allowed value of the neutrino-mass eigen value at 
$M_Z$ decreses slowly 
with increase in the see-saw scale . We obtain the lower bound to be 
 $0.27$ ~eV - $0.16$ ~eV for $M_R= 10^{11}-2\times 10^{18}$ ~GeV.   
\section{V. DEGENERATE NEUTRINOS FROM TYPE II SEESAW AND A MODEL FOR
APPROXIMATE MIXING UNIFICATION}
\label{sec5}

In this section, we address the question of how a quasi-degenerate
neutrino spectrum can arise within a gauge model that
employs the seesaw mechanism for understanding neutrino masses \cite{ref26}.

To begin the discussion, let us present the different forms of the seesaw
mechanism that provide a natural way to understand the small neutrino
masses. Following literature, we will call the two types of seesaw
mechanism as type I and type II. In the type I seesaw mechanism the
neutrino mass matrix is given by the formula, 
\be
M_{\nu}~=~-M_D(fv_R)^{-1}M^T_D
\label{typeI}
\ee
where $f$ is the Majorana Yukawa coupling of the RH neutrinos, $v_R$ is
the $B-L$ symmetry breaking scale, and $M_D$ is the Dirac neutrino-mass 
matrix. In models where information about the
B-L symmetry is not given explicitly, $fv_R$ is replaced by the mass
matrix of the right handed (RH) neutrinos $M_N~=~ fv_R$. Since one expects
the pattern of $M_D$ to be similar to the quark and lepton mass matrices,
one expects the eigenvalues of $M_N$ to be hierarchical and mixing angles
to be small. The Eq.(\ref{typeI}) then tells us that the neutrino masses are
hierarchical. Clearly in such models the radiative magnification of mixing
angles does not occur via the renormalization group evolution as is clear
from Eqs.(3)-(5) in the previous section.

The type I seesaw formula is generic to models which do not have any
connection between
the left and right handed fermions such as in models where one extends the
standard model by adding a right handed neutrino and mass terms for the
RH neutrinos. Things however undergo a drastic
change in models that have asymptotic parity invariance. In such models
there are always Higgs fields that are parity partners of the RH Higgs
fields which give mass to the RH neutrinos. Thus there are operators
which give direct mass to the left handed neutrinos at the same time
that the right handed neutrinos get mass. It turns out also that
the direct neutrino mass term is seesaw suppressed i.e. as the $v_R$ scale
goes to infinity, this contribution, like the right handed neutrino
contribution, vanishes. This direct mass contribution leads to a
modification
of the seesaw formula to the following form (type II seesaw formula
\cite{ref27})
\begin{eqnarray}
M~=~fv_L-M_D(fv_R)^{-1}M^T_D
\label{typeII}
\end{eqnarray}
 Examples  of models where type II seesaw formula arises are left-right
symmetric models or SO(10) models with either $B-L=2$ triplet Higgs fields
or $~B-L=1$ doublet Higgs fields breaking the B-L symmetry. Below we give
an example of a model with triplet Higgs fields. It is important to note
that the renormalization group equations hold for both the type I as
well type II seesaw formula.

The Yukawa coupling matrix $f$ in Eq.(11) that contributes to the 
first term in the seesaw formula as well as the right handed
neutrino mass matrix depends on high scale physics
and is therefore unconstrained by 
by standard model results. We could therefore choose $f$ to be close to
the unit matrix. In this case, quark-lepton unification requires that the
lepton mixing angles be very close to the quark mixing angles but the
neutrino mass spectrum dominated by the first term in Eq.(11) in
combination with second term can easily
lead to a quasi-degenerate spectrum of Majorana neutrinos as well as
approximate mixing unification. In such
schemes, radiative magnification works to provide an understanding of the
large neutrino mixings. The question is whether there is some underlying 
symmetry of the theory for which one can write down a 
natural gauge
model where $f={\bf 1}f_0$ as well as the near unification of quark and
lepton mixings.
Below we provide an example of this kind of model. An
important point is that the renormalization group equations hold for this
type II seesaw formula as long as we assume that the $SU(2)_L$ triplet
Higgs whose vev responsible for the first term in Eq.(11) is heavier than
the seesaw scale. This is true in models realizing the type II seesaw.

We consider a nonsupersymmetric $SU(2)_L\times SU(2)_R\times SU(4)_{PS}$
gauge model with an $S_4$ global symmetry \cite{ref28}. Before describing the
model, a few words about $S_4$ symmetry may be helpful. This
is a nonabelian discrete symmetry group with 24 elements and has the
irreducible representations ${\bf 3, 3', 2, 1', 1}$. We will assign
fundamental fermions to the {\bf 3} dimensional representation of $S_4$
and the Higgs fields $\phi_a$ and B-L=2 triplet fields to representations
of $S_4$ as follows:
$$\br{|c||c|}\hline
{\rm Fields} & S_4\,{\rm rep.} \\ \hline
\Psi_{L,R} (2,1,4)+(1,2,\bar{4}) & {\bf 3} \\ \hline
\phi_0 (2,2,1) & {\bf 1} \\ \hline
\phi_{1,2}(2,2,1) & {\bf 2} \\ \hline
\phi'_{1,2,3}(2,2,1) & {\bf 3} \\ \hline
\Delta_{L,R} (3,1,10)+(1,3,\bar{10}) & {\bf 1} \\ \hline
\er\nonumber$$
$$\Psi=\pmatrix{u_1& u_2&u_3&\nu\cr d_1&d_2&d_3&e}.$$
Let us now write down the $S_4$ invariant  Yukawa couplings:
\par\noindent
\ba 
{\cal L}_Y&=&f_0 (\sum_a\psi^T_{L,a}\Psi_{L,a}\Delta_L + L\leftrightarrow
R)\nonumber\\ 
&&+ h_0\phi_0(\sum_a\bar{\psi}_{L,a}\Psi_{R,a})  
+h_2[(\bar{\psi}_{L,3}\Psi_{R,2}\nonumber\\
&&+\bar{\psi}_{L,2}\Psi_{R,3})\phi_1  
+(\bar{\psi}_{L,3}\Psi_{R,3}+\bar{\psi}_{L,2}\Psi_{R,2}\nonumber\\
&&-
2\bar{\psi}_{L,1}\Psi_{R,1})\phi_2] 
+h_3 [(\bar{\psi}_{L,1}\Psi_{R,3}+\bar{\psi}_{L,3}\Psi_{R,1}\phi'_1\nonumber\\
&&+(\bar{\psi}_{L,2}\Psi_{R,1}+\bar{\psi}_{L,1}\Psi_{R,2})\phi'_2+
(\bar{\psi}_{L,3}\Psi_{R,3}\nonumber\\
&&-\bar{\psi}_{L,2}\Psi_{R,2})\phi'_3]+ h.c.
\label{eq12}\ea
\par\noindent
To get the desired form of the seesaw formula, first note that
$<\Delta^0_L>= v_L\equiv v^2_{wk}/v_R$,  $\Delta^0_R=v_R$, the bidoublet 
vevs are of the form
$$<\phi_i>=\pmatrix{\kappa_i & 0\cr 0 & \kappa'_i}, $$ 
and that $f_0$ is the
identity matrix.

   One can break the $S_4$ symmetry softly so
that all the the $\phi$'s have different vevs. Also note that $h_i$'s can
be complex. Thus six $\phi$'s with independent vevs give us 12 parameters
which is enough to fit the quark mixings and will predict all lepton
mixings equal to quark mixings at the GUT scale. At the GUT scale, this
would predict $m_b=m_{\tau}$ and $m_s=m_\mu$. For the b-quark, this is the
well known $b-\tau$ unification. Using the PDG values for $m_{b,s}$, we
can run it upto the
GUT scale to get $m_b(M_R)\simeq 0.98-1.10$ GeV whereas the corresponding
value of $m_{\tau}\sim 1.18$. However we have for $m_s(M_R)\simeq0.03$
GeV if we use the PDG values. This is about 3 times smaller than the muon
mass at the seesaw scale \cite{ref18}. So we have to add some terms that 
break quark lepton symmetry. 

To cure the $m_s-m_\mu$ problem, we invoke higher dimensional terms
and add a new Higgs multiplet $\Sigma(1,1,15)$ that transforms as
(1,1,15) under $G_{224}$. Also let us assume that $\Sigma(1,1,15)$
transforms like a {\bf 3} dimensional representation of $S_4$ with
only $<\Sigma_3>\neq 0$.The higher dimensional operators that involve
$\Sigma$ have the form 
$\frac{\phi_0}{M} 
[(\bar{\psi}_{L,1}\Psi_{R,3}+\bar{\psi}_{L,3}\Psi_{R,1}\Sigma_1
+(\bar{\psi}_{L,2}\Psi_{R,1}+\bar{\psi}_{L,1}\Psi_{R,2})\Sigma_2+
(\bar{\psi}_{L,3}\Psi_{R,3}-\bar{\psi}_{L,2}\Psi_{R,2})\Sigma_3] $.
Since $\Sigma (1,1,15)$ has a vev
that breaks only $SU(4)_c$ symmetry, it gives different masses to
quarks and leptons. For $<\Sigma_3>/M\simeq 10^{-3}$, this has the right
order of magnitude to lead to the difference between $m_s$ and $m_\mu$ and
not effect the off diagonal elements that are responsible for
mixings. Since the mixing angles go roughly like
$\frac{M_{23}}{M_{33}-M_{22}}$, they do not
 deviate too much from the symmetric values (since $M_{22}\ll M_{33}$).

 As far as the $m_e$ and $m_d$
goes, we can again add nonrenormalizable Yukawa
couplings such as $\bar{\Psi}_L\Psi_R\Delta^{\dagger}_R\Delta_R\phi$ type
terms which will only modify the first generation masses since their
magnitude is of order $v^2_R/M_{P\ell}^2$ down compared to the
renormalizable terms. Again this contribution being a purely diagonal
contribution will change the mixing angles only slightly. Therefore, we
can get a 
 a model of the type we are considering with degenerate neutrinos
and with quark and neutrino mixing angles approximately equal at the 
seesaw scale. This
model can easily be supersymmetrized and all our conclusions go through.

Coming to the neutrino sector, we will first show how type II seesaw
emerges in this model. The complete Higgs content of this model for the
supersymmetric case is: $\Psi (2,1,4); \Psi^c(1,2,\bar{4}), \phi_0
(2,2,1), \phi_{1,2}(2,2,1), \phi'_{1,2,3}(2,2,1),\\
\Delta (3,1,10)\oplus\bar{\Delta}(3,1,\bar{10})$ and $\Delta^c
(1,3,\bar{10})\oplus \overline{\Delta}^c(1,3,10)$ as shown in Table in this
section. In addition we add a Higgs field transforming as $\Omega(3,3,1)$. 
The Higgs part of the superpotential can be written as
\begin{eqnarray}
W'~=~\lambda \Omega
({\Delta}\Delta^c+\overline{\Delta}\overline{\Delta^c}+ Tr{\phi^2_0}+
\cdot\cdot\cdot)
\end{eqnarray}
where $\cdot\cdot\cdot$ denote the $S_4$ singlet bilinears involving the
other $\phi$ fields. Clearly, when we set $F_{\Omega}=0$ to maintain
supersymmetry down to the weak scale, we find that 
$<\Delta^0>\neq 0$. This leads to the type II seesaw which is the cornerstone 
of our discussion.

The gauge group $SU(2)_L\times SU(2)_R \times SU(4)_C$($=G_{224}$) 
is a subgroup
of a numbetr of GUTs like $SO(10)$, $SO(18)$, and $E_6$ etc.
It also contains the subgroups like $SU(2)_L \times SU(2)_R \times U(1)_{B-L}
\times SU(3)_C$($=G_{2213}$)   and the standard model. Thus the model worked 
out  
with $S4 \times G_{224}$   is 
equivalent to a number of underlying high-scale models such as 
$S4\times SO(10)$, $S4 \times SO(18)$, $S4 \times E_6$ etc. It also suggests 
the possibility 
of having $S4\times G_{2213}$ as an approximate symmetry  for quasi-degeneracy.
   
In the absence of such symmetries as dicussed in this section
where a non-abelian discrete symmetry $S4$ occurs along with the gauge 
symmetry $G_{224}$,
~high-scale  unification of quark and neutrino mixings with quasi-degenerate 
neutrinos but with hierachial quark masses  would have been accidental. 
But the type II 
seesaw mechanism in the presence 
of $S4 \times G_{224}$
and its spontaneous breaking guarantees  quasi-degenerate
neutrinos with almost equal mixings in the quark and lepton sectors 
at the high scale while the model fits all the masses and mixings 
at low energies.

\section{VI. CONCLUSION}
\label{sec6}
 In summary, we have shown that in the MSSM, the hypothesis of quark-
lepton mixing unification at the seesaw scale seems to
generate the correct observed mixing pattern for neutrinos i.e. two large
mixings needed for $\nu_e-\nu_\mu$
and  $\nu_\mu-\nu_\tau$ and small mixing for $U_{e3}$ at low energies.
Quasi-degenerate neutrino spectrum with a common mass for neutrinos $\geq
0.1$ eV is a testable prediction of the model. Important new result of our
analysis is that although magnification occurs for the $U_{e3}$ parameter,
it remains small due to the fact that $V_{ub}$ is very small. The
prediction for $U_{e3}$ also provides another test of the model.

Througout this paper we have treated all phases(Majorana and Dirac) to be
vanishingly small in the MNS matrix. It would be interesting to investigate
the effect of phases \cite{ref29} on the implications of our  
mixing unification hypothesis.  
\begin{table*}
\caption{Radiative magnification to bilarge
mixings
at low energies for input values of $\sin\theta_{23}^0
=0.038$,~$\sin\theta_{13}^0=0.0025$, ~and $\sin\theta_{12}^0=0.22$ 
at the see-saw scale $M_R= 10^{13}$ GeV. }
\begin{ruledtabular}
\begin{tabular}{lccccc}\hline
$m_1^0$(eV)&0.3682&0.5170&0.6168&0.7160&0.8160\\
$m_2^0$(eV)&0.3700&0.5200&0.6200&0.7200&0.8200\\
$m_3^0$(eV)&0.4210&0.5910&0.7050&0.8190&0.9330\\
$m_1$(eV)&0.2201&0.3107&0.3719&0.4317&0.4920\\
$m_2$(eV)&0.2223&0.3122&0.3723&0.4324&0.4926\\
$m_3$(eV)&0.2244&0.3152&0.3759&0.0.4366&0.4973\\
$\Dmott$(eV$^2$)&$1.2\times 10^{-4}$&$3.0\times 10^{-4}$&
$3.5\times 10^{-4}$&$6.0\times 10^{-4}$&$5.9\times 10^{-4}$\\
$\Dmttht$(eV$^2$)&$1.0\times 10^{-3}$&$1.8\times 10^{-3}$&
$2.6\times 10^{-3}$&$3.6\times 10^{-3}$&$4.6\times 10^{-3}$\\
$\sin\theta_{23}$&0.667&0.708&0.690&0.677&0.668\\
$\sin\theta_{13}$&0.090&0.104&0.097&0.096&0.090\\
$\sin\theta_{12}$&0.606&0.520&0.604&0.486&0.606\\\hline
\end{tabular}
\end{ruledtabular}
\label{tab1}
\end{table*}
\begin{table*}
\caption{Same as Table I but for different mass scales}
\begin{ruledtabular}
\begin{tabular}{lccc}\hline
{\large \bf $M_R$(GeV)}&{\large \bf $10^{11}$}&{\large\bf $10^{15}$}&
{\large\bf $2\times 10^{18}$}\\
$m_1^0$(eV)&0.4083&0.3970&0.5150\\
$m_2^0$(eV)&0.4100&0.400&0.5200\\
$m_3^0$(eV)&0.4510&0.4730&0.668\\
$m_1$(eV)&0.2723&0.2093&0.1714\\
$m_2$(eV)&0.2726&0.2098&0.1718\\
$m_3$(eV)&0.2745&0.2124&0.1750\\
$\Dmott$(eV$^2$)&$1.6\times 10^{-4}$&$2.0\times 10^{-4}$&
$1.36\times 10^{-4}$\\
$\Dmttht$(eV$^2$)&$1.0\times 10^{-3}$&$1.1\times 10^{-3}$&
$1.1\times 10^{-3}$\\
$\sin\theta_{23}$&0.711&0.682&0.684\\
$\sin\theta_{13}$&0.103&0.098&0.094\\
$\sin\theta_{12}$&0.571&0.463&0.422\\\hline
\end{tabular}
\end{ruledtabular}
\label{tab2} 
\end{table*}

\begin{acknowledgments}
M.K.P. thanks the Institute of Mathematical Sciences
for Senior Associateship.
The work of R.N.M is supported by the NSF grant No.~PHY-0099544. The work
of M.K.P is supported by the DST project No.~SP/S2/K-30/98 of the
Govt.~of India. The work of G.R is supported by the DAE-BRNS Senior
Scientist Scheme
of the Govt of India.
\end{acknowledgments}

\end{document}